\documentclass[nofootinbib,twocolumn,pre,superscriptaddress]{revtex4-1}

\makeatletter
\newif\if@restonecol
\makeatother

\usepackage{graphicx}
\usepackage{amsmath}
\usepackage{hyperref}

\usepackage{newfloat,algcompatible}
\usepackage[size=small]{caption}
\usepackage{etoolbox}

\AtEndEnvironment{algorithm}{\noindent\hrulefill\par\nobreak\vskip-5pt}

\usepackage{newfloat}
\DeclareFloatingEnvironment[
    fileext=loa,
    listname=List of Algorithms,
    name=ALGORITHM,
    placement=tbhp,
]{algorithm}

\DeclareCaptionFormat{algorithms}{\vskip-15pt\hrulefill\par#1#2#3\vskip-6pt\hrulefill}
\algblock[Input]{Input}{EndInput}
\algblockdefx[Input]{Input}{EndInput}%
    [1]{\textbf{Input} #1}%
    {}

\begin{document}

\title{Detecting change points in the large-scale structure of evolving networks}
\author{Leto Peel}
\email{leto.peel@colorado.edu}
\affiliation{Department of Computer Science, University of Colorado, Boulder, CO 80309}
\author{Aaron Clauset}
\email{aaron.clauset@colorado.edu}
\affiliation{Department of Computer Science, University of Colorado, Boulder, CO 80309}
\affiliation{BioFrontiers Institute, University of Colorado, Boulder, CO 80303}
\affiliation{Santa Fe Institute, 1399 Hyde Park Rd., Santa Fe, NM 87501}

\begin{abstract}
Interactions among people or objects are often dynamic in nature and can be represented as a sequence of networks, each providing a snapshot of the interactions over a brief period of time.  
An important task in analyzing such evolving networks is \textit{change-point detection}, in which we both identify the times at which the large-scale pattern of interactions 
changes fundamentally and quantify how large and what kind of change occurred. 
Here, we formalize for the first time the network change-point detection problem within an online probabilistic learning framework and introduce a method that can reliably solve it. 
This method combines a generalized hierarchical random graph model with a Bayesian hypothesis test to quantitatively determine if, when, and precisely how a change point has occurred. We analyze the detectability of our method using synthetic data with known change points of different types and magnitudes, and show that this method is more accurate than several previously used alternatives. Applied to two high-resolution evolving social networks, this method identifies a sequence of change points that align with known external ``shocks'' to these networks.
\end{abstract}

\maketitle

\section{Introduction}
Networks are frequently used as a general framework to quantify and analyze the interactions between objects or people.  Network models can be used to better understand the large-scale structure of interactions by identifying clusters of highly interacting communities or functional groups of structurally equivalent nodes. However, these interactions are often dynamic in nature, and traditional approaches can overlook the non-stationary structure of real networks. In these dynamic and temporally evolving systems we are not only interested in understanding the large-scale structure but also identifying if, when and how it changes in time.  

For instance, in social networks, change points may be the result of normal periodic behavior, as in the weekly transition from weekdays to weekends. In other cases, change points may result from the collective anticipation of or response to external events or system ``shocks''.  Detecting such changes in social networks could provide a better understanding of patterns of social life and an early detection of social stress caused by, e.g, natural or man-made disasters.

Here we define the \textit{network change-point detection} problem and introduce an online probabilistic learning algorithm for solving it.  
Identifying a change point requires inferring a structural ``norm'' for interactions across a sequence of graphs and accurately detecting if, when and how this norm has shifted at some point in time. To characterize what kind and how large a change occurred, we prefer interpretable models of network structure, so that changes in parameter values have direct meaning with respect to the network's large-scale structure.  Here, we take the novel approach of characterizing network norms via probabilistic distributions over graphs, which we learn in an online fashion.
Identifying the timing and shape of such change points divides a network's evolution into contiguous periods of relative structural stability, allowing us to subsequently analyze each period independently, while also facilitating hypotheses about the underlying processes shaping the data.

\begin{figure}
  \begin{center}
    \includegraphics[width=\columnwidth]{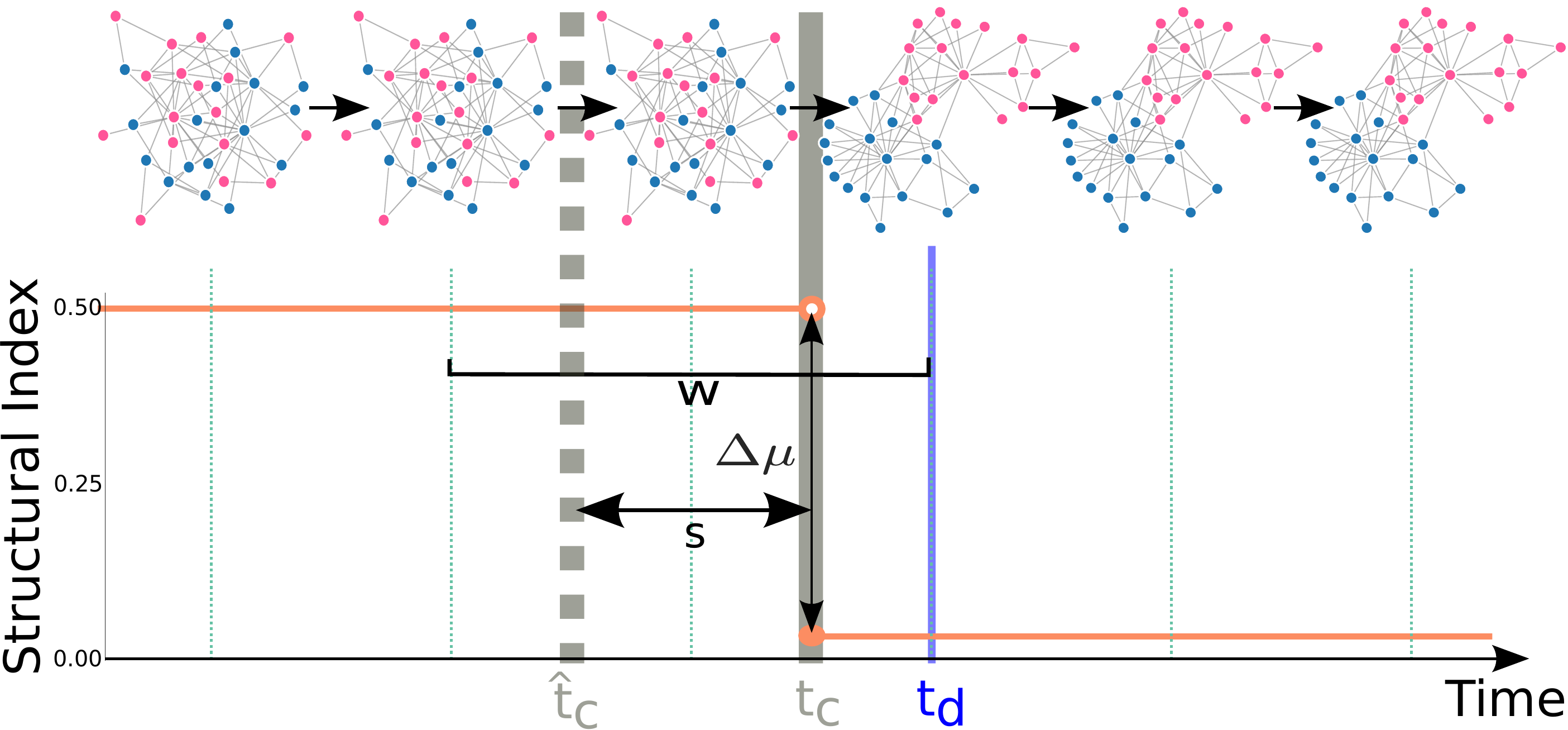}
    \caption{Schematic of a network change point. A sequence of networks in which vertices divide into two groups at time $t_c$, represented by a change in an abstract structural index $\Delta \mu$. To detect this change point, we estimate the time of change $\widehat{t}_c$ within a sliding window of the last $w$ networks, and call $t_d$ the time of detection in which $\widehat{t}_c$ is found to be statistically significant.}
    \label{fig:change_point}
  \end{center}
\end{figure}

Beyond specialized solutions for change-point detection in cybersecurity domains~\cite{levy2009detection}, many ``change-point detection'' methods for networks are in fact solving the distinct anomaly detection problem of identifying network snapshots that deviate significantly from a stationary network norm~\cite{Hiroseetal2009,AkogluFaloutsos2010}. In contrast, traditional change-point detection focuses on the more difficult problem of identifying significant shifts in the norm itself. Solving this problem requires distinguishing a statistically significant change from mere noise, and thus qualitative approaches are likely to be insufficient~\cite{Sunetal2007,berlingerio2013evolving}.  
 
Much like traditional online change-point detection methods for scalar- or vector-valued time series~\cite{BassevilleNikiforov1993}, our approach for the network change-point detection problem has three components:
\begin{enumerate}
  \item select a parametric family of probability distributions appropriate for the data, and a sliding window size $w$;
  \item infer two versions of the model, one representing a change of parameters at a particular point in time within the window, and the one representing the null hypothesis of no change over the entire window; and,
  \item conduct a statistical hypothesis test to choose which model, change or no-change, is the better fit.
\end{enumerate}

Past work on the network change-point detection problem has typically converted the sequence of networks into a time series of scalar values and then applied traditional techniques~\cite{Priebeetal2005,McCullohCarley2011}. 
Here, we introduce a novel solution based on generative models of networks, which define a parametric probability distribution over graphs. 
Our particular choice of model is the generalized hierarchical random graph (GHRG), which compactly models nested community structure at all scales in a network and provides an interpretable output for later analysis. Our approach, however, is entirely general, and the GHRG could be replaced with another generative network model, e.g., stochastic block models~\cite{Hollandetal83,NowickiSnijders2001}, hierarchical graph models~\cite{ClausetMooreNewman2007,BlundellTeh2013}, or Kronecker product graph model~\cite{Leskovecetal2005}. Finally, to choose between the change versus no-change models, we use a Bayesian hypothesis test, with a user-defined parameter specifying a target false-positive rate.

We then show that this approach quantitatively and accurately determines \textit{if} the network norm has changed, \textit{when} precisely the norm change occurred, and \textit{how} the norm has changed. Specifically, we present a taxonomy of different types and sizes of network change points and a quantitative characterization of the difficulty of detecting them using synthetic network data with known change points. We then test the method on two real, high-resolution evolving social networks of physical and digital interactions, showing that it more accurately recovers the timing of known significant external events than comparable techniques.

\section{Defining a probability distribution over networks}
\begin{figure}
  \begin{center}
    \includegraphics[width=\columnwidth]{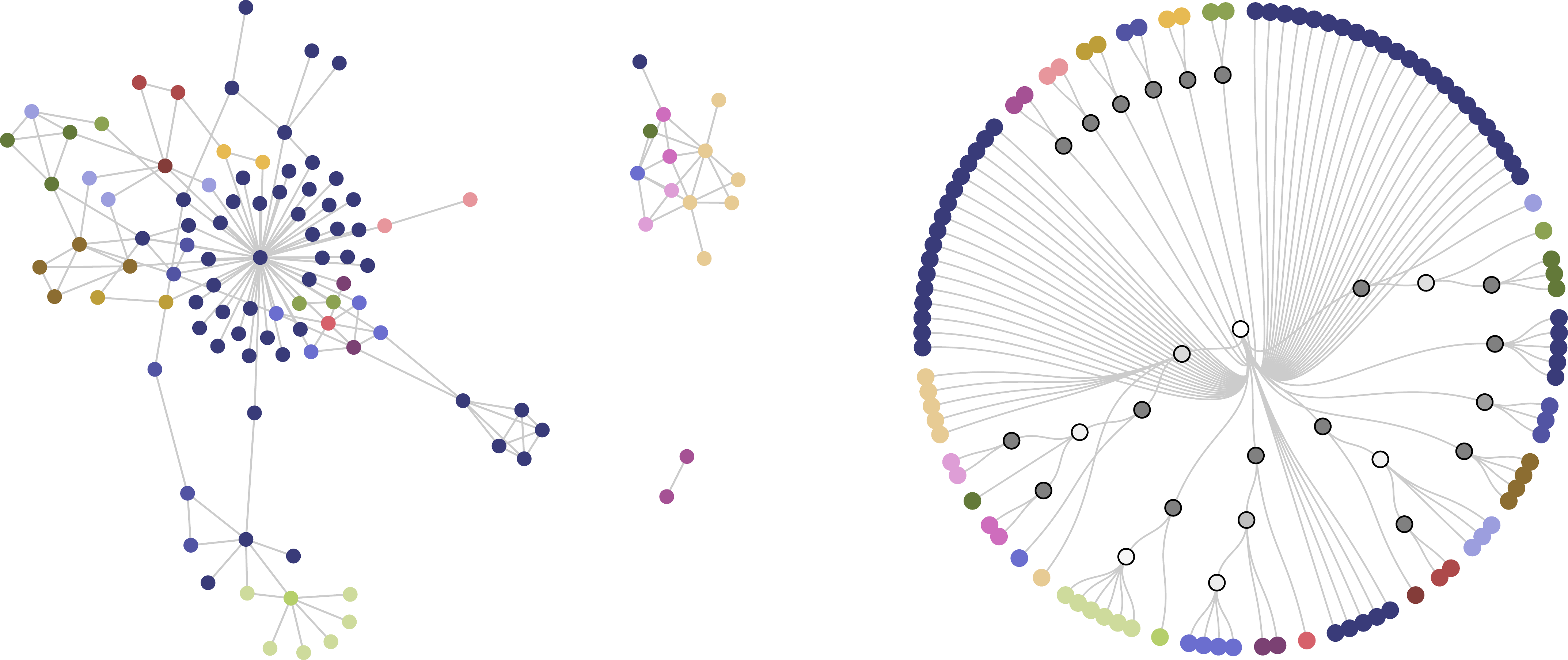}
    \caption{A snapshot of the Enron email network from October 2001 and its corresponding GHRG dendrogram. In the dendrogram, leaves are vertices in the email network and the tree gives their nested group structure.}
    \label{fig:exampleGHRG}
  \end{center}
\end{figure}

Under a probabilistic approach to change-point detection, we must choose a parametric distribution over networks. Here, we introduce the generalized hierarchical random graph (GHRG) model. This model has several features that make it attractive for change-point detection and generalizes the popular hierarchical random graph (HRG) model~\cite{ClausetMooreNewman2007}. First, the GHRG naturally captures both assortative and disassortative community structure patterns, models community structure at all scales in the network, and provides accurate and interpretable fits to social, biological and ecological networks. Second, our generalization relaxes the requirement that the dendrogram is a full binary tree, thereby eliminating the HRG's non-identifiability and  improving the model's interpretability for quantifying how a network's structure varies across a change point.
Third, we use a Bayesian model of connection probabilities that quantifies our uncertainty about the network's underlying generative model.

The GHRG models a network $G=\{V,E\}$ composed of vertices $V$ and edges $E \subseteq \{V \times V\}$. The model decomposes the $N$ vertices into a series of nested groups, whose relationships are represented by 
a dendrogram $T$. Vertices in $G$ are leaves of $T$, and the probability that two vertices $u$ and $v$ connect in $G$ is given by a parameter $p_{r}$ located at their lowest common ancestor in $T$. 
In the classic HRG model~\cite{ClausetMooreNewman2007}, each tree node in $T$ has exactly two subtrees, and $p_{r}$ gives the density of connections 
between the vertices in the left and right subtrees. As a result, distinct combinations of dendrograms and probabilities produce identical distributions over networks, producing a non-identifiable model. In the GHRG, we eliminate this possibility by allowing tree nodes to have any number of children and preferring more compact trees.  Figure~\ref{fig:exampleGHRG} 
illustrates the GHRG applied to a network of email communications.

Given tree $T$ and set of connection probabilities $\{p_{r}\}$, the GHRG defines a distribution over networks and a likelihood function
\begin{equation}
    p(G\,|\,T,\{p_r\}) = \prod_r{p_r^{E_r}(1-p_r)^{N_r-E_r}} \enspace ,
  \label{eq:likelihood}
  \end{equation}
where $E_r$ is the number of edges between vertices with common ancestor $r$ and $N_r$ is the total number of possible edges between vertices with common ancestor $r$:
\begin{equation}
  N_r = \sum\nolimits_{c_i<c_j \in C_r}{|c_i||c_j|}  \enspace ,
\end{equation}  
where $C_r$ is the set of direct descendants of $r$ and $c_i$ is the set of network vertices decending from dendrogram node $i$.

One approach to setting the connection probability parameters $\{p_{r}\}$ would be to choose their values via maximum likelihood, setting each $\hat{p}_r = E_r \left/ N_r \right.$ However, this choice provides little room for uncertainty and is likely to increase our error rate in change-point detection. Consider the case where
exactly zero connections $E_{r}=0$, or equivalently all connections $E_{r}=N_{r}$, are observed for a particular branch $r$. Under maximum likelihood, we set $p_r=0$ or $1$. If a subsequent network has, or lacks, even a single edge whose common ancestor is $r$, then $E_r>0$ or $E_{r}<1$, and the likelihood given by Eq.~\eqref{eq:likelihood} drops to $0$, an unhelpful outcome.

We mitigate this behavior by assuming Bayesian priors on the $p_{r}$ values. Now, instead of setting $p_r$ to a point value, we model each $p_{r}$ as a distribution, which quantifies our uncertainty in its value and prevents its expected value from becoming 0 or 1. For convenience, we employ a Beta distribution with hyperparameters $\alpha=\beta=1$, which corresponds to a uniform distribution over the parameters $p_r$.  Because the Beta distribution is conjugate with the Binomial distribution, we may integrate out each of the $p_r$ parameters analytically 
\begin{equation}
  p(G\,|\,T,\alpha,\beta) \!=\!\prod_r{\frac{\Gamma(\alpha\!+\!\beta)}{\Gamma(\alpha)\Gamma(\beta)}\frac{\Gamma(E_r\!+\!\alpha)\Gamma(N_r\!-\!E_r\!+\!\beta)}{\Gamma(N_r\!+\!\alpha\!+\!\beta)}}  \enspace .
  \label{eq:marginal}
\end{equation}

\begin{figure*}
  \centering
    \includegraphics[width=\textwidth]{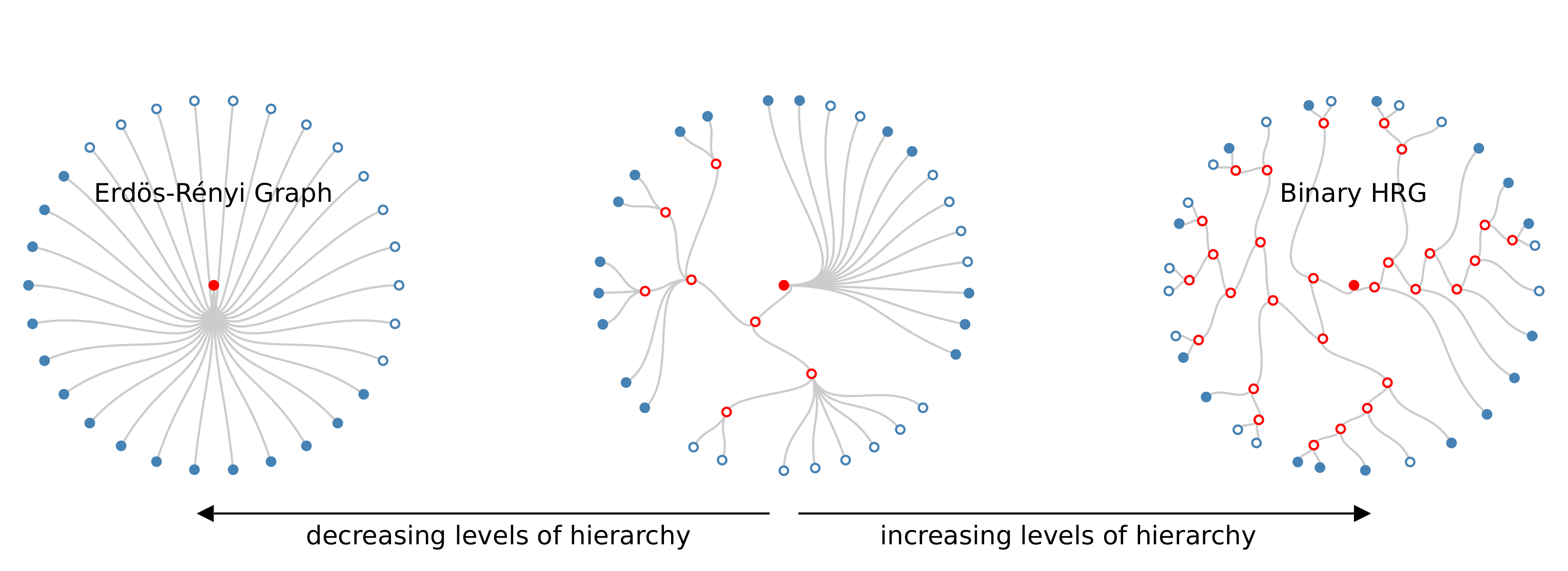}
    \caption{A spectrum of large-scale structure, corresponding to different amounts of hierarchy in the GHRG, ranging from a simple random graph to a complete hierarchical organization.}
    \label{fig:gHRGscale}
\end{figure*}

By relaxing the binary-tree requirement, the GHRG produces a spectrum of hierarchical structure (Fig.~\ref{fig:gHRGscale}). On one end of this spectrum, $T$ contains only a single internal tree node---the root---and every pair of vertices connects with the same probability $p_{r}$ associated with it, equivalent to the popular Erd\H{o}s-R\'enyi random graph model. As more tree nodes and their parameters are added to $T$, the number of levels of hierarchy increases, allowing the model to capture more varied large-scale patterns. In the limit, $T$ is a full binary tree, and we recover the classic HRG model~\cite{ClausetMooreNewman2007}.

\section{Learning the model}
Fitting the GHRG model to a network requires a search over all trees on $N$ leaves and the corresponding link probability sets $\{p_{r}\}$, which we accomplish 
using Bayesian posterior inference and techniques from phylogenetic tree reconstruction.

Tree structures are not amenable to classic convex optimization techniques, and instead must be searched explicitly. However, searching over all non-binary trees is costly. Phylogenetic tree reconstruction faces a similar problem, which is commonly solved by taking a majority ``consensus'' of a set of sampled binary trees~\cite{Bryant2003}. This consensus procedure selects the set of bipartitions on the leaves that occur in a majority of the sampled binary trees, and each such set denotes a unique non-binary tree containing exactly those divisions. For instance, if every sampled tree is identical, then each is identical to the consensus tree; if every sampled tree is a distinct set of bipartitions, the consensus tree has a single internal node to which all of the leaf nodes are connected.
Thus, we estimate $T$ in the GHRG by using a Markov Chain Monte Carlo (MCMC) procedure to first sample 
the posterior distribution of bipartitions. From this set of sampled 
bipartitions, we derive their non-binary majority consensus tree (an approach previously outlined in Ref.~\cite{ClausetMooreNewman2007}, but not used to produce a probabilistic model) and assign link probabilities $\{p_{r}\}$ to the remaining tree nodes.

Given this formulation, we may update the posterior distribution over the parameter $p_r$ given a sequence of observed networks $\{ G_t \}$ by updating the hyperparameters as
\begin{align}
    \tilde{\alpha}_r & = \alpha+\sum_{\{G_t\}}{E_r^{G_t}} \label{eq:alphapost}\\
    \tilde{\beta}_r & = \beta+\sum_{\{G_t\}}{N_r-E_r^{G_t}} \enspace .  \label{eq:betapost}
\end{align}
Thus, we obtain the posterior hyperparameters from the sum of the prior pseudocounts of edges and the empirically observed edge counts (number of present and absent connections).  This Bayesian approach produces an implicit regularization. As the number of observations $N_r$ increases, the posterior distribution becomes increasingly peaked, reflecting a decrease in parameter uncertainty. In the GHRG model, parameters closer to the root of $T$ represent larger-scale structures in $G$ and govern the likelihood of more edges. These parameters are thus estimated with greater certainty, while the distribution over parameters far from the root, representing small-scale structures, have greater variance. This implicit regularization prevents over-fitting to small-scale structural variations and improves the inferred norm's robustness to noise.

\section{Detecting change points in networks}
The final piece of our online network change-point detection method is to determine whether and when the parameters of our current model of ``normal'' connectivity have changed. To accomplish this, we use the posterior Bayes factor over a sliding window of fixed length $w$ to detect if any changes have occurred with respect to a GHRG model fitted over the window. A change is detected if the factor exceeds a threshold determined by a desired false positive rate.

\subsection{Posterior Bayes factor}
To determine whether or not we believe a change has occured within a particular window we use the \textit{posterior Bayes factor}~\cite{Aitkin1991}.  Similar to a likelihood ratio test~\cite{Aitkin1997}, but consistent with our Bayesian framework, the posterior Bayes factor is a ratio of the observed data's likelihood under two different models: a null hypothesis model $H_0$, in which no change occurs, and an alternative hypothesis model $H_1$, in which a change occurs at some particular time $t_{c}$.  However, rather than evaluate the likelihoods under maximium likelihood parameters, we use the posterior marginal likelihood by weighting the average likelihood by the posterior distribution.  For the GHRG, this is calculated by updating the prior hyperparameters $(\alpha,\beta)$ in Eq.~\eqref{eq:marginal} with the posterior hyperparameters $(\widehat{\alpha},\widehat{\beta})$ in Eq.~\eqref{eq:betapost}. 

We restrict our consideration of change points to those within a sliding window of $w$ networks, the last of which is at the ``current'' time $\tau$.  We assume the change point $t_c$ occurs between some pair of snapshots, which we indicate using a 0.5 offset. For the no-change model, we say that all networks within the window were drawn from a model with parameters $\psi^{(\emptyset)}$. For the change model, we let $\psi^{(0)}$ denote the model parameters for networks up to $t_c$ within our window and $\psi^{(1)}$ the parameters for networks after $t_c$, but still within the window. Rewriting the change and no-change hypotheses in terms of such a shift in a parametric distribution over graphs at $t_{c}$, we have
\begin{align}
  H_0 &:  \psi^{(\emptyset)} = \psi^{(\emptyset)} \qquad \textrm{(no change)} \nonumber  \\ 
  H_1 &:  \psi^{(0)} \neq \psi^{(1)} \qquad \textrm{(change)} \enspace .  \nonumber 
\end{align}

Using $\widetilde{\psi} = \{\tilde{\alpha}_r, \tilde{\beta}_r\}$ to denote the set of posterior hyperparameters, the GHRG's posterior Bayes factor for a sequence of graphs $\{G_{\tau-w+1},...,G_{\tau}\}$ is
\begin{align}
  \Lambda_{\widehat{t}_c} = & \sum_{t=\tau-w+1}^{\widehat{t}_c-0.5}{\log p(G_t\,|\,T_{\tau},\widetilde{\psi}_{\widehat{t}_c}^{(0)})} \nonumber \\ 
  & + \sum_{t=\widehat{t}_c+0.5}^{\tau}{\log p(G_t\,|\,T_{\tau},\widetilde{\psi}_{\widehat{t}_c}^{(1)})} \nonumber \\
 & - \sum_{t=\tau-w+1}^{\tau}{\log p(G_t\,|\,T_{\tau},\widetilde{\psi}^{(\emptyset)})} \enspace , \nonumber
\end{align}
where $\widetilde{\psi}^{(\emptyset)}$ is the set of posterior hyperparameters pertaining to the no-change hypothesis (no change point anywhere in the window of $w$ networks), while $\widetilde{\psi}_{\widehat{t}_c}^{(0)}$  and $\widetilde{\psi}_{\widehat{t}_c}^{(1)}$ are the hyperparameters for the networks up to and then following the point $\widehat{t}_c$, respectively.

Finally, the time at which the change occurs $t_{c}$ is itself an unknown value, and must be estimated.  
We make the conservative choice, choosing $\widehat{t}_c$ as the time point between a pair of consecutive networks that maximizes our test statistic $\Lambda$ across the window. Letting $g_{\tau}$ for a given window of $w$ networks ending at $\tau$ be that largest value
\begin{equation}
  g_{\tau} = \max_{\tau-w+1 \,<\, \widehat{t}_c \,<\, \tau} \Lambda_{\widehat{t}_c} \enspace ,
  \label{eq:decision}
\end{equation}
we then say that the time of detection $t_{d}$ is the first time point $\tau$ at which $g_{\tau}$ exceeds a threshold $h$ (called a ``stopping rule'' in the change-point detection literature):
\begin{equation}
  t_d = \min \{\tau : g_{\tau} > h \} \enspace .
\end{equation}

\begin{figure*}
  \begin{center}
    \includegraphics[width=\textwidth]{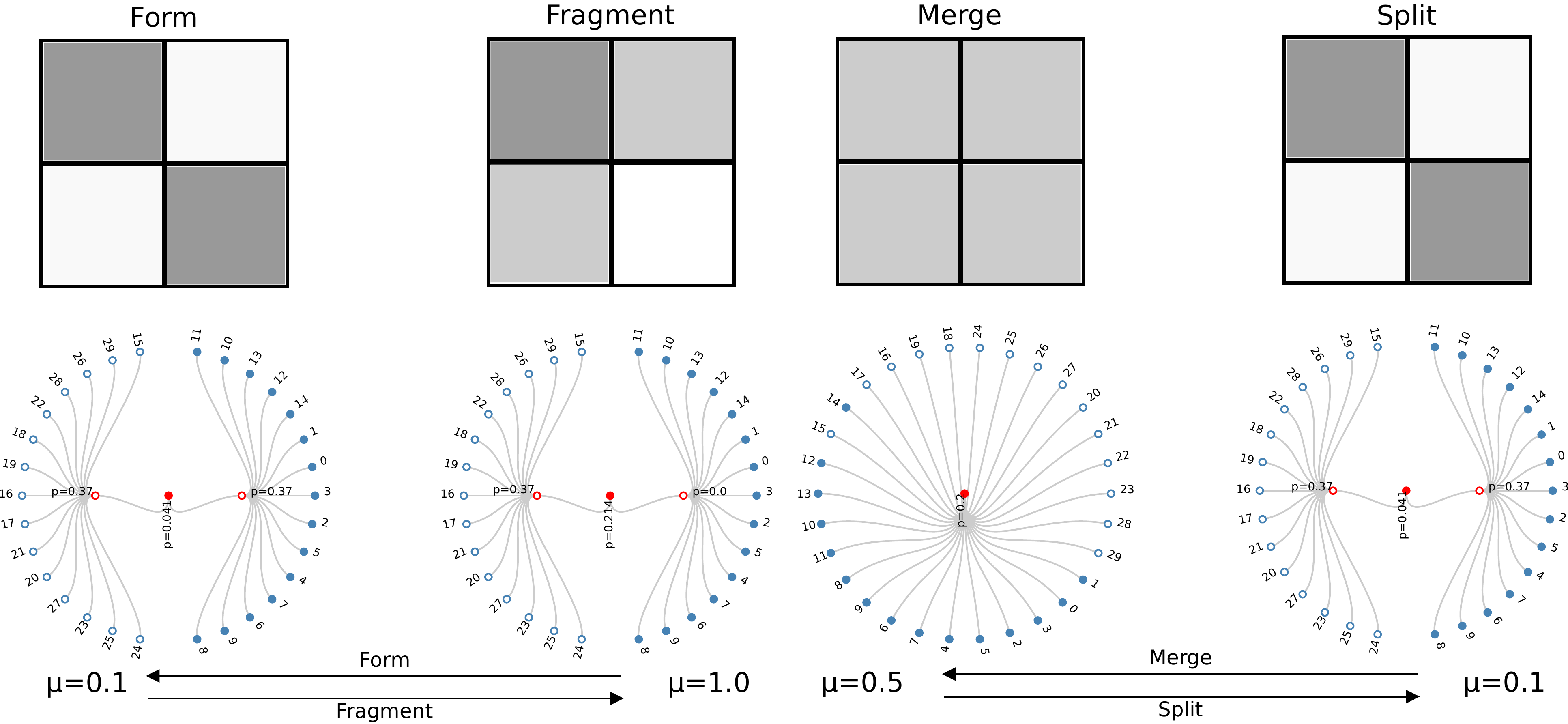}
    \caption{Taxonomy of change points: formation versus fragmentation and merging versus splitting. In each case, we show both the block structure of the adjacency matrix for two groups and the corresponding GHRG model. For our experiments, we switch from one structure to the other by changing the structural index $\mu$. See text for more detail.}    
    \label{fig:synthcartoon}
  \end{center}
\end{figure*}

\subsection{Parametric bootstrapping}
The choice of threshold $h$, which $g_{\tau}$ must exceed for a detection to occur, sets 
the method's resulting false positive rate and the distribution of $g_{\tau}$ under the null model.  
Recent results on model comparison for statistical models of networks, and specifically the stochastic block model, of which the GHRG is a particularly useful special case, suggest that for technical reasons the null distribution can deviate substantially from the $\chi^2$ distribution~\cite{Yanetal2012}. To avoid a misspecified test, we estimate the null distribution numerically, via Monte Carlo samples from a parametric bootstrap distribution~\cite{EfronTibshirani1993} defined by the GHRG for the no-change model. In this way, we estimate the null distribution exactly, rather than via a possibly misspecified approximation.
  
For each network we sample from the no-change GHRG model and calculate $g_{\tau}$ from Eq.~\eqref{eq:decision} to obtain its distribution under the hypothesis of no change (see Algorithm \ref{alg:bootstrap}).
  Using the sampled distribution, the threshold $h$ may then be chosen so that $p(g_{\tau} > h) = p_{\rm{fp}}$ is the desired false positive rate.  In practice we do this by calculating a $p$-value for the test case by counting the proportion of likelihood ratios in our null distribution that are higher than our test statistic $g_{\tau}$:
\begin{equation}
  p\text{-value} = \frac{|\{g_{\tau}\}_{\rm{null}} > g_{\tau}|}{|\{g_{\tau}\}_{\rm{null}}|} \enspace .
\end{equation}
Thus, if we find a $p$-value below the chosen threshold, we say a change is detected, and when the no-change model is correct, we are incorrect no more than $p_{\rm{fp}}$ of the time.
 
\begin{algorithm}[t]
 \caption{Parametric bootstrap sampling $g_{\tau}$ from the null model distribution}
 \label{alg:bootstrap}
 \begin{algorithmic}[H]
 \Input $\mathcal{G} = \{G\}_{\tau-w+1}^{\tau}$ \EndInput
 \STATE $\{g_{\tau}\}_{\rm{null}} = \emptyset.$
 \STATE GHRG$(T_{\tau},\widetilde{\psi}_{\tau}) =$ fitGHRG($\mathcal{G}$).
 \FOR{$i = 1$ to $1000$}
 \STATE sample $w$ graphs, $\{G_t\}_{t=1}^{w} \sim$ GHRG$(T_{\tau},\widetilde{\psi}_{\tau})$.\;
 \FOR{$\tilde{t}_c = 1$ to $w$}
 \STATE calculate $\widetilde{\psi}^{\emptyset},\widetilde{\psi}^{0}, \widetilde{\psi}^{1}$ according to Eq.~\eqref{eq:alphapost} and Eq.~\eqref{eq:betapost}.\;
 \STATE $\Lambda_{\tilde{t}_c} = \sum_{t=1}^{\tilde{t}_c-1}{\log p\left(G_t|T_{\tau},\widetilde{\psi}_{\tilde{t}_c}^{(0)}\right)} $\\ $\qquad + \sum_{t=\tilde{t}_c}^{w}{\log p\left(G_t|T_{\tau},\widetilde{\psi}_{\tilde{t}_c}^{(1)}\right)}$\\ $\qquad - \sum_{t=1}^{w}{\log p\left(G_t|T_{\tau},\widetilde{\psi}^{\emptyset}\right)}$ \;
 \ENDFOR
 \STATE $g_{\tau}^{(i)} = \max_{\tilde{t}_c} \Lambda_{\tilde{t}_c}$.\;
 \STATE $\{g_{\tau}\}_{\rm{null}} = \{g_{\tau}\}_{\rm{null}} + g_{\tau}^{(i)}$.\;
 \ENDFOR
\end{algorithmic}
\end{algorithm}

\begin{figure*}[t]
  \centering
    \includegraphics[width=\textwidth]{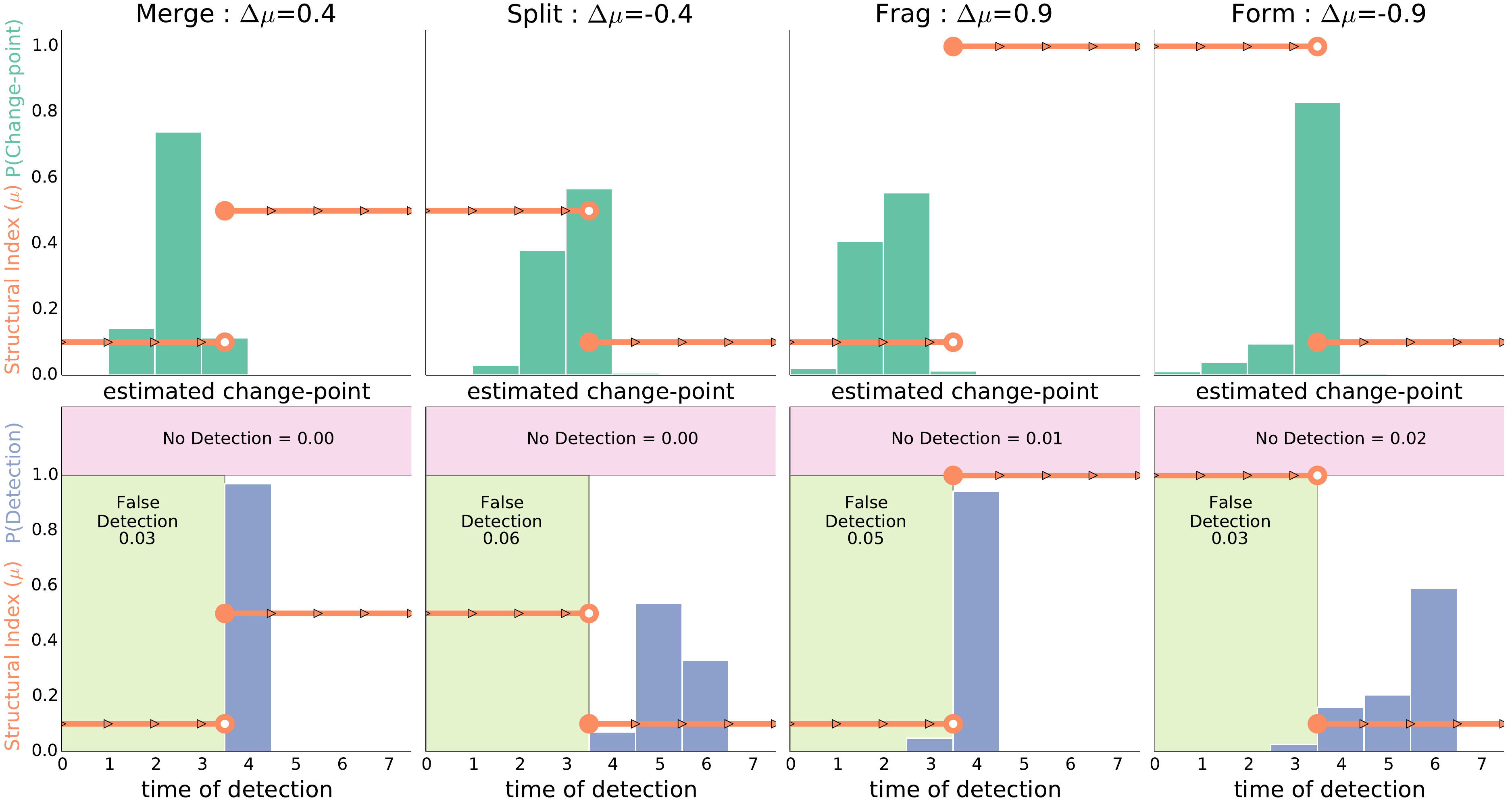}
    \caption{Results for merge, split, fragment and form change points, where the size of the change is parameterized by the structural index $\mu$ (orange line) and a discontinuity occurs at the true point of change. The resulting distributions of (\textit{top row}) estimated change points times $\widehat{t}_{c}$ (green bars), and (\textit{bottom row}) detection times $t_{d}$ (blue bars), for each change type. The 0.5 time offset indicates that the estimated change point time $\widehat{t}_{c}$ occurs between a pair of consecutive network snapshots, while $t_d$ gives the time of the last network in the window when a detection event occurs.  False detections (false positives) occur when $t_{d}<t_{c}$; no detection (false negatives) occur when $\tau-w+1>t_{c}$ without a detection.}
    \label{fig:synth_distribution}
\end{figure*}

\section{Detectability of change points}
Before applying our method to empirical data with unknown structure and unknown change points, we first systematically characterize the detectability of different types of network change points under controlled circumstances on synthetic data, generated using our GHRG model, with known structure and changes.  

The following change-point types constitute difficult but realistic tests that cover a broad variety of empirically observed large-scale changes to network structure. For our numerical tests, we choose network with $N\!=\!30$ vertices and a sparse and constant expected number of connections (marginal link probability of $0.2$). These small networks provide a more difficult test case than larger networks because the limited amount of data make change points harder to detect since it becomes harder to characterize the norms. 
Furthermore, we define four general types of change points: \textit{splitting}, when one large community divides in two; \textit{merging}, when two communities combine (the time-reversal of splitting); \textit{formation}, when one of two groups of vertices add edges to make a community; and \textit{fragmentation}, when one of two groups loses all its edges (the time-reversal of formation).

Defining the structural index 
$\mu = p_{\rm out} \left/ (p_{\rm in}+p_{\rm out})\right .$ 
provides a single parameter that controls the switching between these distinct states.
For the merge/split change points, we choose the merged state to be $\mu=0.5$, which produces a single community in which every edge occurs with the same probability $p_{\rm{in}}=p_{\rm{out}}$. In the split state, the network is comprised of two distinct communities. 

For formation/fragmentation change points, we use the same two-community model, but now fix the link probability within one community and use $\mu$ to describe relationship between the $p_{\rm{in}}$ and $p_{\rm{out}}$ of the second community.

We now summarize these change points with respect to $\mu$:
\begin{align}
  &\rm{merge}\!\! &\mu \neq 0.5 &\rightarrow \mu=0.5& \quad\!\! &\rm{fragment} \!\! &\mu < 1 &\rightarrow \mu=1 \notag\\
  &\rm{split} &\mu = 0.5 &\rightarrow \mu \neq 0.5& \quad\!\! &\rm{form}  &\mu = 1 &\rightarrow \mu < 1 \notag \enspace .
\end{align}
All tests used a $w=4$ window size and a 0.05 false-positive rate. For comparison, we used three simpler versions of our probabilistic framework in which we replace the generative model with a univariate Gaussian and convert the network sequence into a time series of scalar values (mean degree, mean geodesic distance, mean local clustering coefficient).

For each of the change types, Figure~\ref{fig:synth_distribution}  shows two different distributions (over 100 runs): the estimated change point $\widehat{t}_c$ and the time of detection $t_d$ (see Figure~\ref{fig:change_point}).  We find that the estimated change points tend to either be correct or slightly early.  The time of detection (the end of the sliding window when a change is detected) quantifies how many networks after the change we must see before we identify the change point. We find that the merge and fragment changes are detected quickly, while their change points are often estimated early. In contrast, the split and formation changes are detected later, while the estimation of the change points themselves is more accurate.

\begin{figure}
  \centering
    \includegraphics[width=\columnwidth]{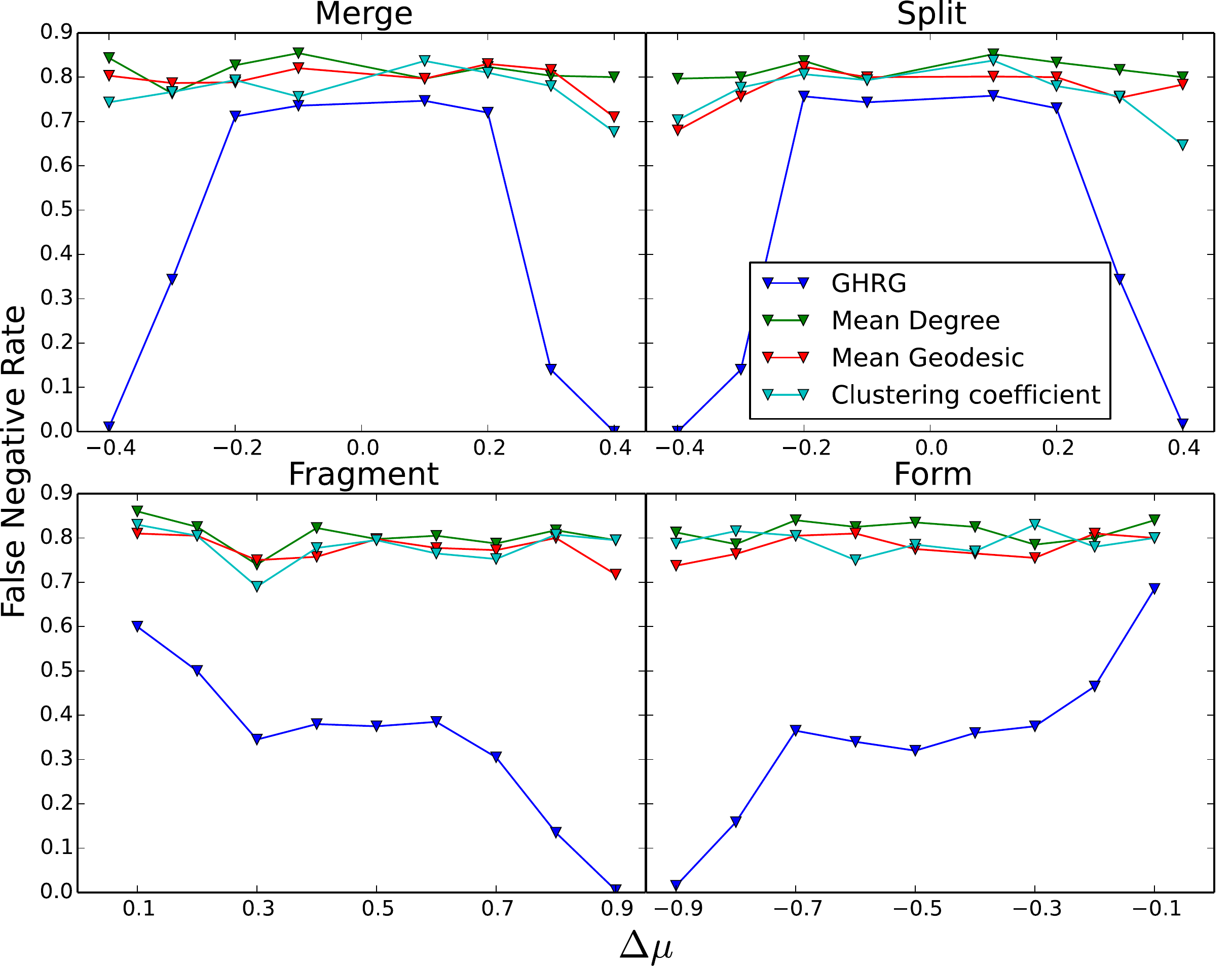}
    \caption{False negative (\textit{bottom}) error rates for our method and the simple methods on the four different change types for different magnitudes of change($\Delta\mu$).}
    \label{fig:synth_results}
\end{figure}

In Figure~\ref{fig:synth_results} we compare the false positive and false negative error rates among all four methods. On false positives, all methods are close to 0.05, which matches the desired false alarm rate. However, the false negative rates differ widely, with the simple methods performing terribly in nearly every case, even when the size of the change is large. In contrast, our method performs well across all four tests, except when the size of the change is very small, e.g., when $\Delta\mu\approx0$, which represents the hardest cases, where small-sample fluctuations obscure much of the actual change.  In the ``split'' and ``merge'' experiments we notice that there seems to be a threshold around which the detectability rapidly changes.  In these experiments, for changes of small magnitude, the networks are close to Erd\H{o}s-R\'enyi random graph both before and after the change point. It is likely then, that this is related to the detectability phase transition known to occur in the community detection problem~\cite{Decelleetal2011}.

\section{Change points in real networks}
We now apply these approaches%
\footnote{Code for our method is available at \url{http://tinyurl.com/letopeel/code.html}} 
to detect changes in two high-resolution evolving networks, the MIT Reality Mining proximity network%
\footnote{\url{realitycommons.media.mit.edu/realitymining.html}}
~\cite{EaglePentland2006} and the Enron email network%
\footnote{\url{www.cs.cmu.edu/~enron/}}
~\cite{KlimtYang2004}, for which a set of external ``shocks'' exists that serve as targets for change-point detection. Both data sets are evolving networks of human social interactions, but represent different interaction types.

We quantify the performance in terms of precision and recall as a function of the detection delay $s$ between estimated change points $\{\widehat{t}_c\}$ and scheduled events $\{t_c\}$, i.e.,  
\begin{align}
  \rm{Precision}(s) &= \frac{1}{n_c}\sum_i{\delta \left(\inf_j\left|\widehat{t}^{(i)}_c-t_c^{(j)}\right| \leq s\right)}  \\
  \rm{Recall}(s) &= \frac{1}{n_a}\sum_j{\delta \left(\inf_i\left|\widehat{t}^{(i)}_c-t_c^{(j)}\right| \leq s\right)},
\end{align}
where $\delta(x)$ is a delta function that equals $1$ if $x$ is true and $0$ otherwise, and $n_c$ and $n_a$ are the number of estimated change points and actual events respectively.  The precision is then the proportion of estimated change points that occur within a given delay $s$ of a known event.  Similarly, recall is the proportion of known events that occur within a delay $s$ of an estimated change point.

\begin{figure}
\centering
    \includegraphics[width=\columnwidth]{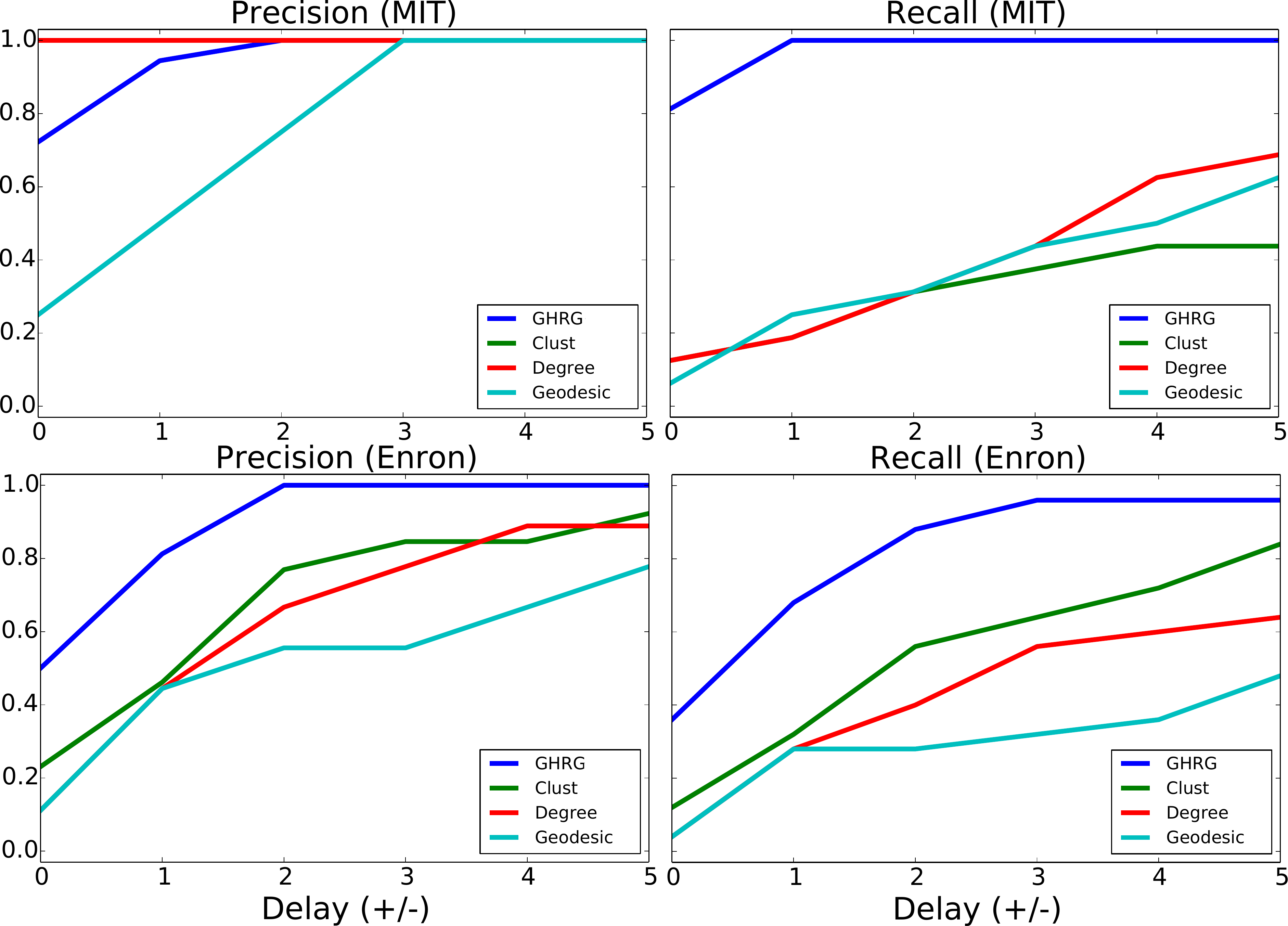}
  \caption{Precision and recall of our method and the three baseline methods at identifying known external events as a function of delay ($s$).}
    \label{fig:real_prec_rec}
\end{figure}

\subsection{Social proximity network}
The MIT network is comprised of proximity data for 97 faculty and graduate students, recorded continuously via Bluetooth scans from their mobile phone over 35 weeks~\cite{EaglePentland2006}. 
From the raw scan data, we extracted a sequence of weekly networks, in which an edge denotes physical proximity to one of the 97 subjects at some point that week.  Associated with the dataset are 16 known external events including public holidays, spring and winter breaks, exam periods, etc~\cite{Eagle2005}.  

For each detection method, the GHRG and the three simple methods, we used a window size $w=4$, the same as in the synthetic experiments. The results for each method are shown in Figure~\ref{fig:real_prec_rec} (top).  We see that clustering coefficient and mean degree slightly outperform our approach in terms of precision, but our approach is much better than all the baseline approaches at recall.  Closer examination of the change-points detected with each of the simple methods reveals that they exhibit low sensitivity relative to the known external events~\cite{Eagle2005}.
In particular, they do pick out several real change points, including Winter break (mean geodesic distance), the beginning (mean degree) and end (clustering coefficient) of the independent activities period. However, they also miss the majority of other events.  In particular mean degree and clustering coefficient only detect a total of 2 change points, which explains the high precision scores.  Furthermore there is little consistency across these methods (with the exception of the beginning of Sponsor week). Thus, these techniques seem both unreliable and inconsistent.

In contrast, the GHRG method identifies nearly all of the known external events, along with a few additional change points, e.g., one week before and one week after Sponsor week. This fact agrees well with the social dynamics of Sponsor week, an event involving 75 of the subjects and which typically shifts work schedules dramatically as they seek to meet deadlines and project goals~\cite{Eagle2005}. 

\begin{figure*}
  \begin{center}
    \includegraphics[width=\textwidth]{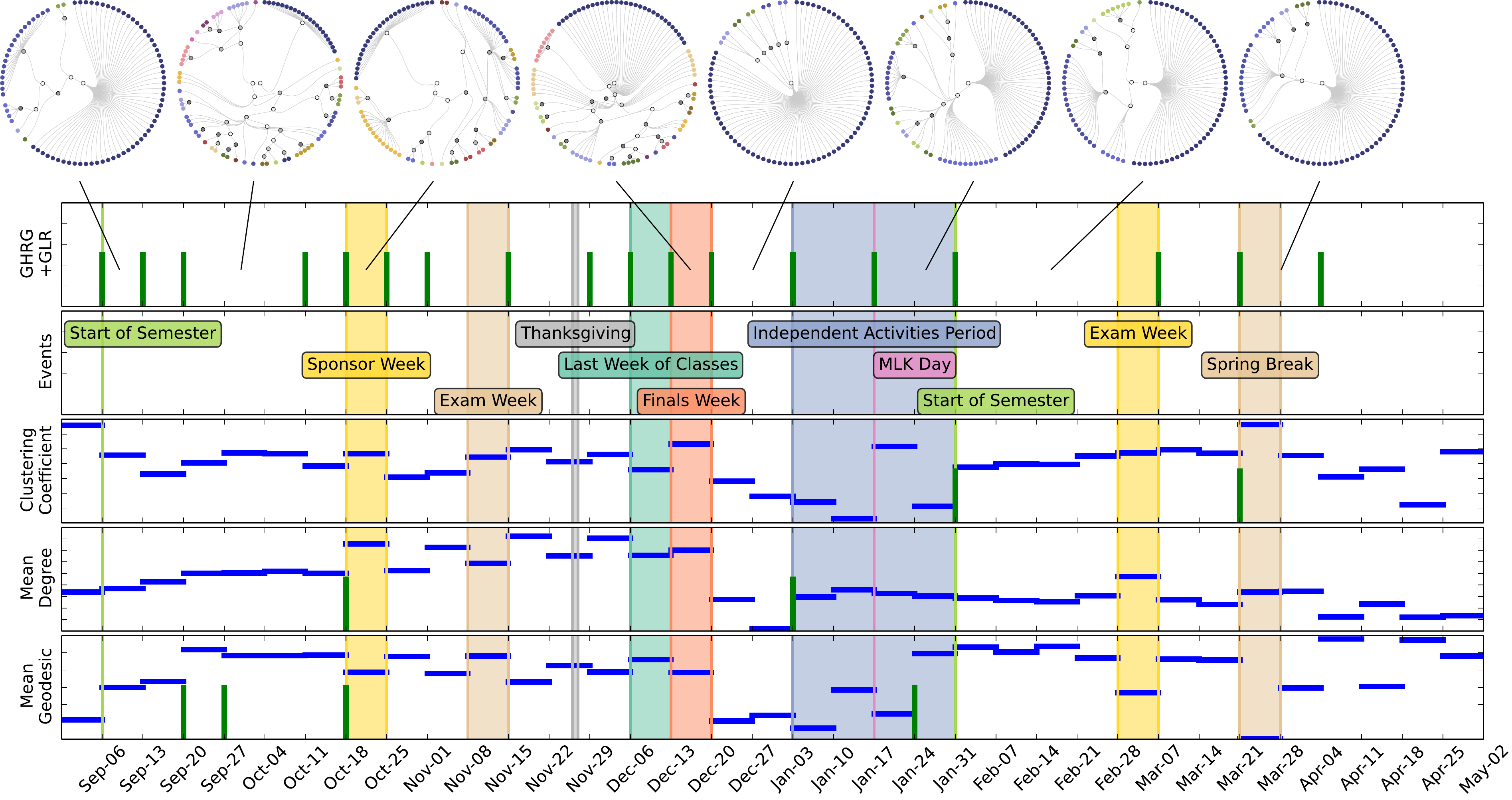}\\ 
    \vspace{15mm}
    \includegraphics[width=\textwidth]{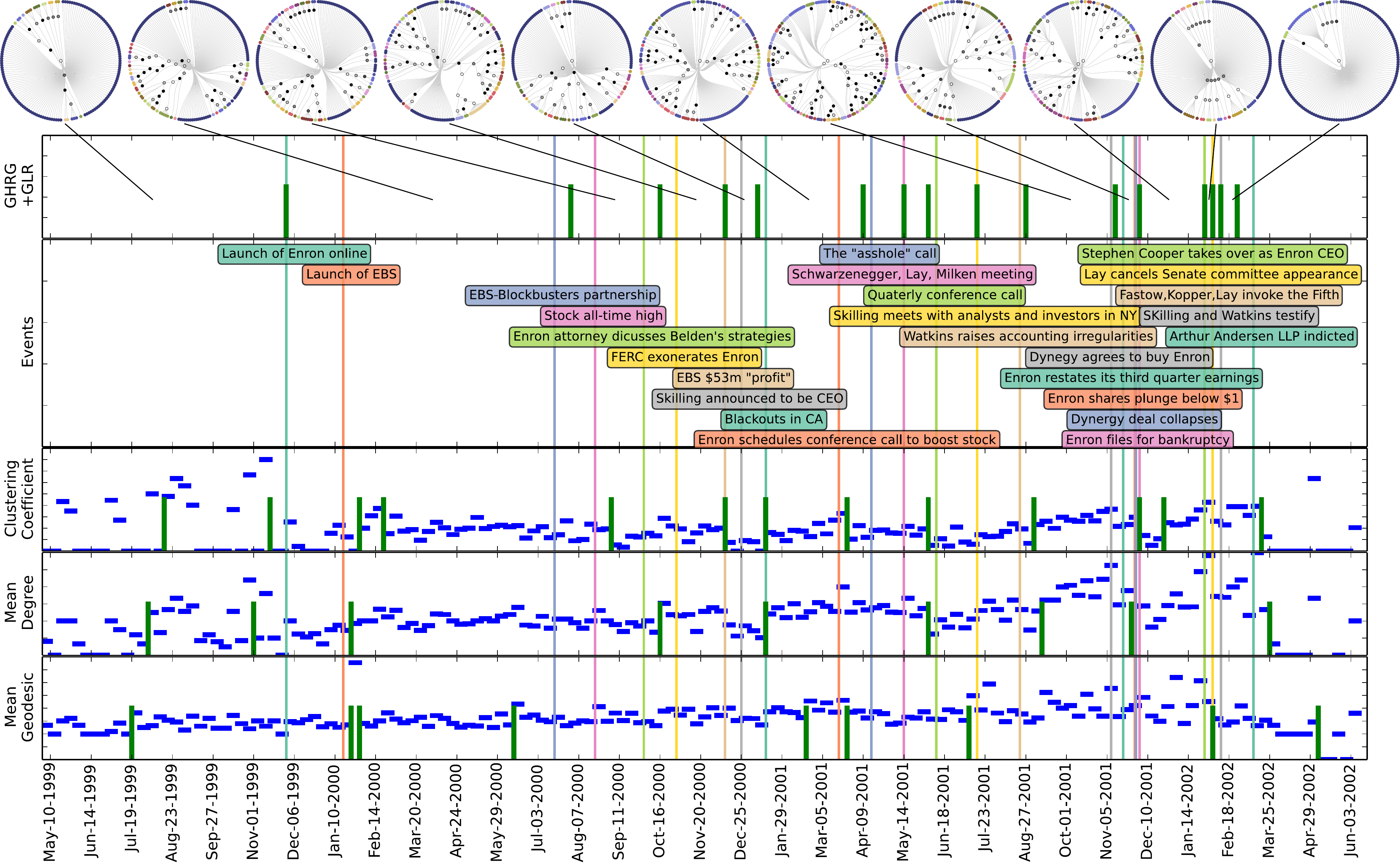}
    \caption{Change points detected in the MIT proximity (\textit{top}) and Enron email (\textit{bottom}) networks.  Each figure shows
     the change points detected (green bars) using our method followed by the three simple methods (along with the values of the relevant summary statistic).  Known events are indicated by the colored vertical bars spanning all methods.}
    \label{fig:real_networks}
  \end{center}
\end{figure*}

Additionally, the GHRG method finds more change points in the Fall semester than in the Spring. Examining the dendrograms themselves, we find that the changes in the inferred structures in the Fall are much more dramatic than in the Spring; see Figure~\ref{fig:real_networks} (top). This agrees with the fact that 35 of the subjects were new students in the Fall semester and thus still establishing their social patterns~\cite{Eagle2005}. By the Spring semester, these patterns had largely stabilized, and the large perturbation of Sponsor week was absent. Overall, the GHRG both recovers known events, highlights additional changes, and provides an interpretable basis for discovering new patterns within this evolving network.

\subsection{Enron email network}
The Enron network is comprised of emails among 151 users, mostly senior management of the Enron energy company.  We identified a total of 25 events during the 3 year timeline. These data were made public by the Federal Energy Regulatory Commission during its investigation after the company's collapse. 
Using a cleaned version of these data~\cite{KlimtYang2004}, we applied both the GHRG and simple methods to weekly snapshots from May 1999 to June 2002. 

Because this network sequence is very long, we examined the impact of varying window size, choosing $w=\{4,8,16\}$. Results across window sizes were highly consistent, although larger values produced additional change points. This suggests that window size may operate like a temporal resolution parameter, with longer windows giving more resolution. The results for each method are shown in Figure~\ref{fig:real_prec_rec} (bottom) using a window size $w=16$.

The results for each method are shown in Figure~\ref{fig:real_networks} (bottom).  As with the MIT network, we again find that the simple methods perform poorly; the GHRG method performs much better than all of them in both precision and recall. 
Examining the GHRG change points and the list of external events, we find that the identified change points correlate well with key meetings and events such share price fluctuations. Examining the inferred dendrograms, we find that a particularly large structural change occurred around the launch of Enron online (Nov 1999).  
Before its launch, the network's structure is very sparse, while after launch, the number of levels in the GHRG model increase dramatically reflecting the formation of many communities.  We also see that the structure and density of the networks increase over the period starting immediately after the Californian blackouts through until Stephen Cooper takes over as CEO and the collapse of the company is imminent.

\section{Discussion}
When analyzing a sequence of time evolving networks, a central goal is to understand how the network's structure has changed over time, and how it might change in the future. Change-point detection provides a principled approach to this problem, by decomposing a potentially non-stationary sequence of networks into subsequences of distinct but probabilistically stationary structural patterns. 
Here, we have presented the first change-point detection method for evolving networks that utilizes generative network models and statistical hypothesis tests.  By formalizing this problem within a probabilistic framework, we developed a statistically principled method that can detect, in an online fashion, if, when and how such change points occur in the large-scale patterns of interactions. Under our framework, change points occur when the shape of an estimated probability distribution over networks changes significantly.

Not all such change points are equally easy to detect. Using synthetic data with known structure and known change points, we observed that only changes of a large enough magnitude could be detected reliably.  Furthermore, we found that changes associated with two communities merging or with one of several communities losing its internal connections (``fragmentation'') were more difficult to accurately detect than those associated with one community splitting in two or with many singletons connecting to form a new community (``formation''). This asymmetry in the detectability of different types of network changes begs the question of whether more sophisticated techniques can eliminate these differences, and whether adding auxiliary information like edge weights \cite{Aicher26062014} or vertex attributes \cite{peel2011topological} makes this problem easier or harder. 

That being said, change-point methods based on network measures like the mean degree, clustering coefficient, or mean geodesic path length performed poorly, yielding high false negative rates even for large structural changes (Fig.~\ref{fig:synth_results}). This poor performance is likely the result of network measures discarding much of the specific information that generative models utilize. Applied to two high-resolution evolving social networks, our method provided very good results, recovering the timing, from network data alone, of many more known external ``shock'' events than the network-measure methods (Fig.~\ref{fig:real_prec_rec}).

\begin{figure}
  \centering
  \includegraphics[width=\columnwidth]{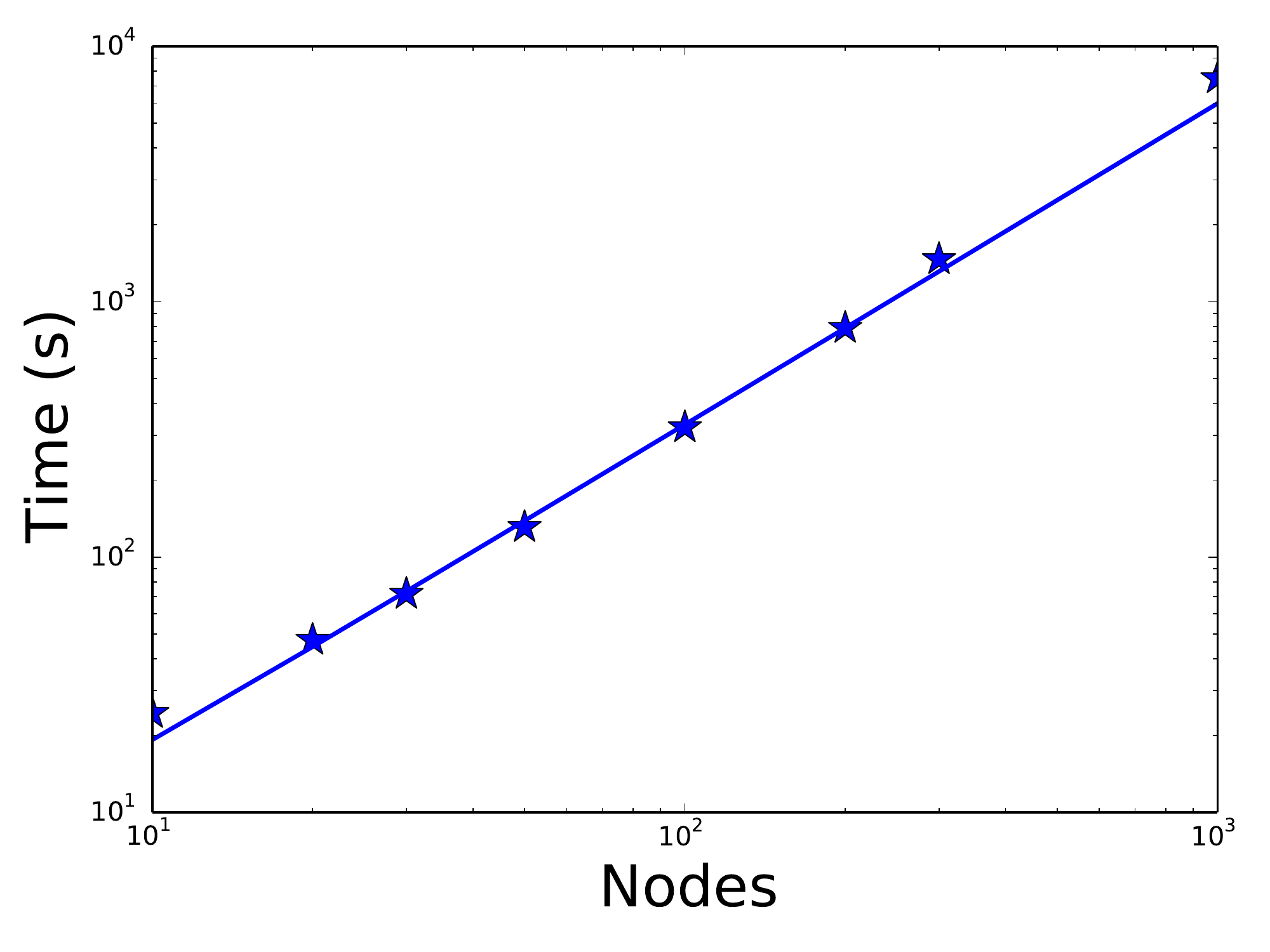}
  \caption{Time taken to infer GHRG for a window of 4 networks of varying size.  Assumes sparse graphs, i.e., O(N) edges.}
  \label{fig:time}
\end{figure}

The computational burden of our current implementation lies in the MCMC procedure used to infer the hierarchical structure (see Figure \ref{fig:time} for inference times).  The recent work of Ref.~\cite{BlundellTeh2013} proposes a greedy approach to inferring a hierarchy that could naturally be used within our change-point detection as a more scalable alternative.  However, it is important to understand what the trade-off is between accuracy and scalability, and we believe that this would be an interesting and useful direction for future work.

Although the GHRG model yielded good results, in principle, any generative model could be used in its place, e.g., the stochastic block model~\cite{Hollandetal83,NowickiSnijders2001,Aicher26062014} or the Kronecker product graph model~\cite{Leskovecetal2005}.  Similarly, the recent work in graph hypothesis testing~\cite{MorenoNeville2013} could potentially be adapted to the change-point detection problem. Two key features of the GHRG model for change-point detection, however, are its interpretability and the way it naturally adapts its dendrogram structure to fit the network, adding or removing levels in the hierarchy, as the network evolves.  
Combined with the strong results on synthetic and real-world data, this approach to change-point detection promises to have broad application, perhaps particularly in social networks, where interpretability provides a crucial bridge to testing hypotheses about the underlying social dynamics driving network evolution.

\section{Acknowledgements}
We thank Dan Larremore for helpful conversations, and acknowledge support from Grant \#FA9550-12-1-0432 from the U.S. Air Force Office of Scientific Research (AFOSR) and the Defense Advanced Research Projects Agency (DARPA).


\end{document}